\documentclass[aps,pra,twocolumn,superscriptaddress]{revtex4-2}
\usepackage{amsfonts}
\usepackage{amsmath}
\usepackage{mathrsfs}
\usepackage{amssymb}
\usepackage{graphicx}
\usepackage{lipsum} 
\usepackage[T1]{fontenc}
\usepackage{epstopdf}
\usepackage{bm}
\usepackage{color}

\usepackage[colorlinks=true,linkcolor=blue,anchorcolor=blue,citecolor=blue,urlcolor=blue]{hyperref}

\newcommand{\bPsi}{{\bm \Psi}}
\newcommand{\bpsi}{{\bm \psi}}
\newcommand{\bvarphi}{{\bm \varphi}}
\newcommand{\bg}{{\bm g}}

\newcommand{\bw}{{\bm w}}
\newcommand{\tmu}{\tilde{\mu}}

\newcommand{\halpha}{\hat{\alpha}}

\newcommand{\rev}[1]{\textcolor{black}{{#1}}}

\begin{document}

\title{Wannier solitons in spin-orbit-coupled Bose-Einstein condensates in optical lattices with a flat-band.}

\author{Chenhui Wang}
\email{knightwch@outlook.com}
\affiliation{International Center of Quantum Artificial Intelligence for Science and Technology (QuArtist) and Department of Physics, Shanghai University, Shanghai 200444, China}
\affiliation{
 Centro de F\'{i}sica Te\'orica e Computacional, Faculdade de Ci\^encias, Universidade de Lisboa, Campo Grande, Ed. C8, Lisboa 1749-016, Portugal 
 }

\author{Yongping Zhang}
\email{yongping11@t.shu.edu.cn}
\affiliation{International Center of Quantum Artificial Intelligence for Science and Technology (QuArtist) and Department of Physics, Shanghai University, Shanghai 200444, China}

\author{V. V. Konotop}
\email{vvkonotop@ciencias.ulisboa.pt}
\affiliation{
 Centro de F\'{i}sica Te\'orica e Computacional, Faculdade de Ci\^encias, Universidade de Lisboa, Campo Grande, Ed. C8, Lisboa 1749-016, Portugal 
 }
 \affiliation{
 Departamento de F\'{i}sica, Faculdade de Ci\^encias, Universidade de Lisboa, Campo Grande, Ed. C8, Lisboa 1749-016, Portugal 
 }

\begin{abstract}

We investigate families of soliton solutions in a spin-orbit coupled Bose-Einstein condensate embedded in an optical lattice, which bifurcate from the nearly flat lowest band. Unlike the conventional gap solitons the obtained solutions have the shape well approximated by a Wannier function (or a few Wannier functions) of the underlying linear Hamiltonian with amplitudes varying along the family and with nearly constant widths. The Wannier solitons (WSs) sharing all symmetries of the system Hamiltonian are found to be stable. Such solutions allow for the construction of Wannier breathers, that can be viewed as nonlinearly coupled one-hump solitons. The breathers are well described by a few-mode model and manifest stable behavior either in an oscillatory regime with balanced average populations or in a self-trapping regime characterized by unbalanced atomic populations of the local potential minima (similarly to the conventional boson Josephson junction), with the frequencies controlled by the inter-atomic interactions.

\end{abstract}

\maketitle

\section{Introduction}
Periodic modulation of parameters of a medium, where a wave propagates, introduces artificial dispersion. If the medium is nonlinear the existence of gap solitons becomes possible. This is a well-known phenomenon thoroughly studied in diverse physical settings including Bose-Einstein condensates (BECs)~\cite{BraKon2004, MorOber2006, Zhang2009} and nonlinear optics~\cite{KarMysTor2009}. Such soliton-bearing systems are described by one or several coupled Gross-Pitaevskii (GP) or nonlinear Schr\"odinger (NLS) equations with periodic potentials. For BECs loaded in optical lattices (OLs), the attractive nonlinearity originated by a negative scattering length of the inter-atomic interactions enables solitons characterized by chemical potentials, $\mu$, located in either a semi-infinite or in a finite gap of the spectrum of the underlying linear Hamiltonian. Otherwise, if the nonlinearity is repulsive (i.e., the scattering length is positive), solitons may exist with chemical potentials belonging only to finite gaps. In either of the cases, gap solitons are localized solutions belonging to families which are usually characterized by the dependence of the number of atoms $N$ ({\it alias} norm) on the chemical potential. In a common situation,  families of solitons bifurcate from one of the edges of a linear band $\mu_{\nu k}$ ($\nu$ is the index of a band and $k$ is the Bloch wavenumber). In the small-amplitude limit, $N\to 0$, such gap solitons are governed by an NLS equation for their envelopes, whose effective dispersion is determined by  $\partial_k^2 \mu_{\nu k}$ (see e.g.~\cite{KonSal2002}). Respectively, when a soliton amplitude tends to zero its width infinitely increases approaching the respective Bloch state it bifurcates from.  

The described situation, however, is not applicable anymore if $\partial_k^2 \mu_{\nu k}=0$. Such a situation is encountered in optical applications, where in the presence of non-zero dispersion, $\partial_k^4 \mu_{\nu k}\neq 0$, and the nonlinearity support so-called quartic solitons~\cite{Sterke2021}, which can also exist in SO-BECs without lattices (as it is discussed below).

It turns out that physical systems can have strongly or even completely suppressed linear dispersion, when several or all derivatives $\partial_k^n \mu_{\nu k}=0$ (for $n=1,2,...$) vanish. The last situation occurs, for instance, in the case of flat bands existing in the spectra of certain discrete systems. In that case, the spatial localization of waves is possible even in the purely linear limit, while in the nonlinear case, it allows for the existence of discrete compactons~\cite{Yulin2013}, unconventional families of nonlinear modes~\cite{Vicencio2013}, formation of solitons with abrupt edges~\cite{Goblot2019}, and unusual expansion regimes~\cite{Leykam2013}.

Being ubiquitous in discrete systems~\cite{Leykam2018, Poblete2021}, flat bands in continuous periodic systems were studied mainly in the tight-binding limit allowing for approximating such systems by discrete lattices. 
The nonlinear effects of genuinely continuous systems featuring one or a few flat bands well separated from the rest of the spectrum did not receive attention yet. Meantime there is an essential subtle difference between the concept of flat bands in discrete systems and continuous systems with periodic potentials. A discrete system bearing an ideally flat band, which is obtained as a tight-binding limit of the respective continuous model features nonzero inter-site coupling which is determined by the Fourier coefficients of the spectrum (see e.g.~\cite{AKKS}). In contrast, nearly perfect flat bands of a continuous system mean zero coefficients of the Fourier expansion, and thus completely decoupled sites, in the tight-binding limit (see the discussion below). Furthermore, the existence of perfectly flat bands in continuous systems is impossible, i.e., the concept of flatness itself must be specified. 

The existence of nearly flat bands is a characteristic feature of continuous spin-orbit-coupled BECs (SO-BECs) with judiciously chosen parameters~\cite{Zhang2013, Zhang2015, Kartashov2016a, Kartashov2016b}. Discrete models describing a SO-BEC within the framework of a tight-binding limit were considered in~\cite{Yan2014, Abdullaev2018, Gligoric2016}. Being created in atomic gases~\cite{Spielman}, SO-BECs remain in the focus of current experimental studies~\cite{experiment1,experiment2}. However, the existence and properties of gap soliton bifurcating from a flat band in continuous models have not been addressed.  

In this work, we are interested in a minimal continuous SO-BEC model with periodic coefficients (described in Sec.~\ref{flat band}) which features a single nearly flat band that is separated by a finite gap from the rest of the spectrum. We obtain numerically and describe analytically (in Sec.~\ref{soliton})  soliton families which in the limit of small amplitudes are states which are well approximated by the Wannier functions (WFs), differing from families bifurcating from the Bloch states of non-flat bands, as it happens for the conventional gap solitons. Such matter wavepackets remain localized even when their norm becomes negligibly small and are termed~\cite{AKKS} as {\em Wannier solitons} (WSs).
Another peculiarity of the WSs (addressed in Sec.~\ref{breather}) is that their enhanced stability allows for the construction of stable multi-hump Wannier breathers. Such excitations alternatively can be viewed as analogs of bosonic Josephson junctions (BJJ) based on two, three, or even more BEC clouds, in which, however, the atomic exchange among different clouds is enabled by inter-atomic interactions, and hence is not observed in the linear regime.  The outcomes are summarized in the Conclusion. 

\section{Flat bands enabled by spin-orbit coupling}
\label{flat band}

We consider a quasi-one-dimensional SO-BEC loaded in an OL~\cite{Hamner2015} which in the mean-field approximation is described by the dimensionless Gross-Pitaevskii (GP) equation for the spinor ${\bPsi}=\left({\Psi}_{1},{\Psi}_{2}\right)^{T}$: 
\begin{align}
	\label{GPdimless}
i\partial_t\bPsi  = H\bPsi+g\left(\bPsi^\dagger\bPsi\right)\bPsi,
\end{align}
where  
\begin{align}
\label{H}
H=-\frac{1}{2} \partial_x^{2}-i\gamma\sigma_{z}\partial_x +\frac{\Omega}{2}\sigma_{x} 
+V_0\sin^{2}x,
\end{align}
 is the linear Hamiltonian,  $\gamma$ is the SOC strength, $\Omega$ is the Rabi frequency, $\sigma_{x,y,z}$ are the  Pauli matrices, $V_{0}$ is the amplitude of the OL and $g$ is the effective atomic interaction.
 The units of the energy and the spatial length are chosen as $2E_{L}$ and $1/k_{L}$, respectively. Here, $E_L=\hbar^{2}k_{L}^{2}/2m$ is the recoil energy of the OL, $m$ is the atom mass, and $k_L$ is the wave number of the lattice beams. The dimensionless order parameter is normalized  
 $N=\int_{-\infty}^{\infty}\bPsi^\dagger\bPsi dx$ with the norm $N$ related to the physical number of atoms $\mathcal{N}$:  $\mathcal{N}=NN_0$, where $N_0=gE_L/(\hbar\omega_{r}k_{L}a_{s})$, $\omega_{r}$ is the trap frequency along the transverse directions, and $a_{s}$ is the s-wave scattering lengths.

Considering the typical experimental data for a $^{87}$Rb BEC from Ref.~\cite{Hamner2015}:   $\omega_r =2\pi\times 150\,$Hz, $a_s=5.29\,$nm,   $k_L=2.88\mu\, {\rm m}^{-1}$, $E_L\approx3.2\times10^{-31}$J, $\Omega=7.62E_L$ and $V_0=-1.4E_L$, we obtain the dimensionless parameters $\gamma=k_R/k_L=1.96$,  with $k_R=5.63\,\mu{\rm m}^{-1}$ being the wave number of the Raman beams, $\Omega=3.81$, $V_0=-0.7$ and $N_0\approx217$. In experiments, $\gamma$ can be varied either by changing the angle between the incident Raman beams or through fast modulation of the intensities of the Raman lasers~\cite{experiment3, Zhang2013_1}. The Rabi frequency $\Omega$ and the amplitude of the OL $V_0$ are free parameters that can be tuned by changing the intensity of Raman and lattice beams, respectively.

It is known~\cite{Zhang2013, Zhang2015, Kartashov2016a, Kartashov2016b} that $\gamma$ and $\Omega$ considered as control parameters can be chosen to ensure the extreme flatness of the lowest band of the linear spectrum $\mu_\nu(k)$ of the Hamiltonian $H$, defined by the eigenvalue problem $H\bvarphi_{\nu k}=\mu_\nu(k)\bvarphi_{\nu k}$, where $\bvarphi_{\nu k}(x)$
$\nu$ is the band index, and $k\in[-1,1)$ is the Bloch wavenumber. Indeed, consider first the case where $\Omega=2\gamma^2\gg 1$. Neglecting the OL in the leading approximation obtain the linear dispersion relation for the lower branch $\mu_-(k)$ of the spectrum: $\mu_-=-\Omega/2\left[1-k^4/(2\Omega^2)+\mathcal{O}(k^6/(2\Omega^3))\right]$. Thus, for any given finite interval of wavenumbers, $k$, one can choose $\Omega$ and $\gamma$ large enough to make as many orders of the dispersion negligible as necessary. In particular, leaving only the leading term $\propto k^4$ in the mentioned expansion one can look for quartic SO-BEC solitons as approximate solutions.  Adding now an OL that restricts the wavenumbers to the first Brillouin zone, should result in the nearly flat lowest band. 

\begin{figure}[t]
	\includegraphics[width=0.5\textwidth,trim={0.4cm 0.8cm 4cm 2.3cm}]{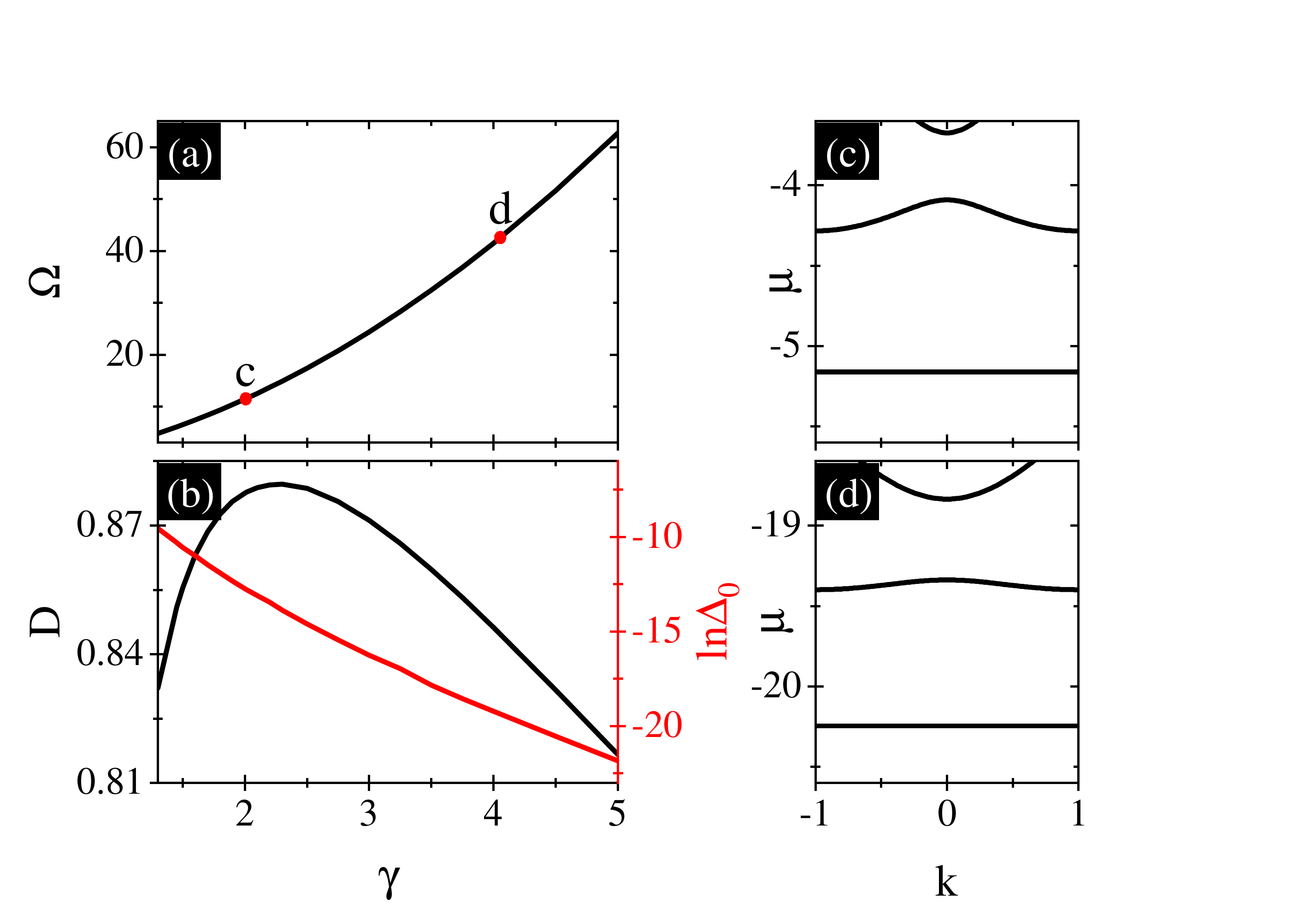}
	\caption{In the Rabi frequency $\Omega$ (a) and the gap width $D$ between the flat lowest and first excited bands (black) and the maximum lowest band flatness $\Delta_{0}$ (red) in (b) {\it versus} the SOC strength $\gamma$ with $V_0=2$. The linear Bloch spectrum is illustrated for the points marked by red dots and the respective letters for \rev{arbitrarily chosen parameters} $\gamma=2.01$, $\Omega=11.4844$ in (c) and $\gamma=4$, $\Omega=41.4649$ in (d).}
        \label{fig:one}
\end{figure}

There is however a limitation to such a straightforward approach. While formally soliton families can bifurcate from such a flat band towards the semi-infinite gap (what happens in the case of attractive condensates), the finite gaps created by a very shallow OL are also negligible, i.e., practically no soliton families in finite gaps can be found. Therefore, we start by briefly describing a situation where a nearly flat band is separated from the rest of the spectrum by a non-vanishing finite gap. While the very existence of such SO-BECs was illustrated in~\cite{Zhang2013, Zhang2015},  Fig.~\ref{fig:one} we show that flat bands can be encountered for a sufficiently wide range of physically accessible parameters. 

To characterize the width of the lowest band $\mu_0(k)$ we represent 
\begin{align}
\label{mu}
    \mu_0(k)=\mu_0+\tmu(k)
\end{align}
where $\mu_0$ without argument stands for $\mu_0(0)$ (recall that $k\in[-1,1)$). In a continuous model, exactly flat bands, i.e., bands with $\tmu(k)\equiv 0$, are impossible. Thus, the flatness of the lowest band ($\nu=0$) implies a smallness of $\tmu (k)$, which in this work, will be quantified by the parameter 
\begin{align}
\label{delta}
  \Delta_0=\max \mu_0(k)-\min\mu_0(k).
\end{align}
 We say that the band is {\em nearly flat} (or {\em flat}, for brevity) if $ \Delta_0\lesssim 10^{-4}$.
 The so-defined criterion is not strict and can be modified. Our choice, in addition to ensuring the smallness of the bandwidth, also aims to optimize several requirements, including the possibility of operating within a sufficiently wide range of parameters (see Fig.~\ref{fig:one} below), to enlarge the region of validity of the estimate $N\propto\mu$, and to squeeze the transition region between this region and the edge of the linear band, holding at the same time a sufficient fidelity of the numerical results.
On the other hand, to characterize the size of the lowest finite gap we introduce the parameter $D=\min \mu_1(k)-\max\mu_0(k)$. 

In Fig.~\ref{fig:one}(a) we illustrate 
the parameter dependence $\Omega(\gamma)$ in the range of SOC strengths where the flat band coexists with a sufficiently large gap [see $\Delta$ (red) and $D$ (black) curves in Fig.~\ref{fig:one}(b)]. The examples of the two lowest bands, panels (c) and (d), illustrate the features of the SO-BEC in a finite-depth lattice (in all calculations, we fix  $V_0=2$) featuring the only flat band, all other hands being non-flat. The flatness of the shown lowest bands in the absence of SOC, i.e., at $\gamma=0$, is $0.17$. Such band-gap structures significantly differ from the typical models of BEC in deep optical lattices with several flat lower bands (see e.g.~\cite{AKKS}). Note also, that the flat lowest band of a SO-BEC without periodic potential would require much higher values of the Rabi frequency (according to the above discussion).

The conventional gap solitons, described by the NLS equation, require either $|\mu_0''(k_0)|\gg |\mu_0^{(m)}(k_0)|$, where $k_0$ is a Bloch wavenumber at which the edge of a finite or semi-infinite gap is achieved.  This allows one to neglect (or to account perturbatively) the higher dispersion. When $\mu_0''(k_0)$ is  comparable with the forth-order dispersion $\mu_0^{(4)}(k_0)$ artificially induced by the OL, one still can obtain gap solitons (such solitons were considered for optical applications~\cite{Cole2014}). A real-world BEC, available in an experiment has a finite extent.  Hence below a certain amplitude, a conventional gap soliton with a wide (exceeding the real size of the condensate) envelope of the respective Bloch state cannot be observed experimentally. The case considered here is the opposite one: in the leading approximation {\em all} orders of the dispersion of the lowest band are negligible, i.e., the {\em approximation of an ideally flat band} is valid. As we will show below, in the leading order widths of such solitons remain finite until very small amplitudes, i.e., very close to the linear limit.

\section{Wannier solitons}
\label{soliton}

\subsection{Small-amplitude expansion.}
\label{subsec:expansion}

Having the Bloch basis given, one can define~\cite{Kohn1959} the two-component WFs. For the lowest flat band, one has 
\begin{align}
\label{Wannier} 
\bw_n(x) =\frac{1}{2}\int_{-1}^{1}\bvarphi_{0 k}(x)e^{-i\pi kn}dk.
\end{align}
where $n$ numbers the minima of the OL.
Applying $H$ to the WF we obtain
\begin{align}
\label{Hw}
		H\bw_{n}=\mu_{0} \bw_{n}(x)+\sum_{m\neq n} f_{ n-m}\bw_{m}(x),
\end{align} 
where  
\begin{align}
    f_m&=\frac{1}{2}\int_{-1}^1\tmu(k)  e^{ikm\pi}dk 
\end{align}
and it is used that due to $\pi$ periodicity $\tmu(k)=\tmu(k+2)$ [see (\ref{mu})]. 
It is relevant to mention that the last term in Eq.~(\ref{Hw}) while having a small amplitude determined by the flatness of the band, is delocalized in space, unlike the first term involving $\bw_n(x)$, which is localized in the vicinity of the points $x_n=n\pi$.

To describe the soliton families bifurcating from the flat band, we recall that all $f_m$ are nearly zero (zero would correspond to an ideally flat band). Thus, assuming (without loss of generality) that a soliton family bifurcates from the linear mode centered at $x=0$, 
we look for a solution of Eq.~(\ref{GPdimless}) in the form of the "mixed" (Wannier-- for the lowest band and Bloch-- for all other bands) expansion 
\begin{align}
\label{expansion}
\Psi\left(x,t\right)=\sqrt{\epsilon} e^{-i\mu t}\left[a\left(t\right)\bw_0\left(x\right)+\epsilon \bpsi_1+\mathcal{O}\left(\epsilon^2\right)
\right],
\end{align}
where $\epsilon\ll 1 $ is a formal small parameter, $\mu=\mu_0+f_0$, 
\begin{align}
\label{expansion_psi}
\bpsi_1&= \sum_{n\neq 0} b_n(t)\bw_n(x)+\sum_{\nu=1}^\infty \int_{-1}^1  b_{\nu k} (t)\bvarphi_{\nu k}(x) dk, 
\end{align}
and only the lowest-order terms are accounted for.  
Substituting expansions (\ref{expansion}) and (\ref{expansion_psi}) in Eq.~(\ref{GPdimless}),  and projecting on the WFs $\bw_0$ and $\bw_n$ (with $n\neq0$) withing the accepted accuracy we obtain respectively:
\begin{align}
\label{eq_a}
    i \partial_t a 
    = \epsilon \chi_0|a|^2a+\epsilon \sum_{n\neq 0}  b_n f_{n},
    \\
 \label{eq_b}   
  \epsilon i \partial_t b_n -\epsilon\sum_{m\neq 0}  b_mf_{m-n}
  = af_{-n}   +\epsilon \chi_n |a|^2a,
 \\
 \label{eq_bk}  
   i\partial_{t} b_{\nu k}+[\mu-\mu_\nu(k)] b_{\nu k} =\chi_{\nu k}|a|^2a,
\end{align}   
where 
\begin{align}
\label{chi_n}
   \chi_n=&g\langle\bw_n,(\bw_0^\dagger,\bw_0)\bw_0\rangle,
    \\
    \chi_{\nu k}=&g\langle\bvarphi_{\nu k},(\bw_0^\dagger,\bw_0)\bw_0\rangle,
\end{align}
are the effective nonlinear coefficients, and we use the notation $\langle {\bm f},\bg\rangle=\int_{-\infty}^{\infty} {\bm f}^\dagger(x)\bg(x)dx$. The solutions described by (\ref{expansion}) belong to the family of WSs. 

The expansion (\ref{expansion}) does not allow for the formal limit $\epsilon\to 0$ because $\bw_0(x)$ is not an eigenstate of the linear Hamiltonian. However, one can show that for an interval of amplitudes, parametrized by a small parameter $\epsilon$: 
\begin{align}
\label{eps}
    \Delta_0\lesssim \epsilon\ll 1,
\end{align}
formula (\ref{expansion}) accurately describes the family of solutions. Indeed, subject to the assumption  $|b_nf_n|\sim |b_n|\Delta_0 \ll |a(t)|$ [it is considered that $a(t)\sim 1$], Eq.~(\ref{eq_a}) is readily solved
\begin{align}
\label{solut:a}
    a(t)=a_0e^{-i\epsilon \chi_0|a_0|^2t},
\end{align}
where $a_0$ is the initial amplitude. Then, in the l.h.s. of Eq.~(\ref{eq_b}) one can neglect the hopping terms and obtain
\begin{align}
    \epsilon b_n\approx \left(\frac{f_{-n}}{\epsilon \chi_0 |a_0|^2}+\frac{\chi_n}{\chi_0} \right) \left(e^{-i\epsilon\chi_0 |a_0|^2t}-1\right). 
\end{align}
Since $|\chi_n|\ll \chi_0$ due to the localization of the WFs, one concludes that limiting inequality $\epsilon |b_n|\ll |a_0|$ remains valid for times $t\ll 1/\Delta_0$ provided (\ref{eps}) is satisfied. Imposing  the condition $\Delta_0\ll \epsilon$ [cf. (\ref{eps})] one obtains that the required smallness of $\epsilon |b_n|$ holds for any time. Finally, we notice that it follows from (\ref{eq_bk}) that $b_{\nu k}$ terms do not display secular growth, and thus give a small correction to the leading term in the expansion (\ref{expansion}).

In the limit $\epsilon\to 0$, or more precisely when $ \epsilon \ll \Delta_0$, expansion (\ref{expansion}) is not valid, because of the secular term in (\ref{eq_b}). To find the soliton family in this limit one should consider the conventional multiple-scale expansion~\cite{KonSal2002,BraKon2004} characterized by the scaling $N \sim\sqrt{|\mu-\mu_0|}\sim\epsilon$. It turns out that such scaling also implies that the width of the soliton is of the order of $1/\epsilon$, which in the case at hand corresponds to $10^{4}$ (recall that according to our definition, the flatness of a band requires $\Delta_0\lesssim 10^{-4}$) or to even more lattice periods in the dimensionless units. Meantime, lengths of experimentally feasible condensates typically are of the order of $500\,\mu m$~\cite{Hamner2015}, which is about $10^{3}$ in the dimensionless units. Thus the extremely tiny "boundary" domain in the vicinity of the flat-band edges, which is excluded from our consideration here, is also beyond the physical accessibility in a real system.

\subsection{Wannier soliton families}
\label{subsec:families}

\begin{figure}[t]
	\includegraphics[ width=0.5\textwidth,trim={1.5cm 1.0cm 0cm 1cm}]{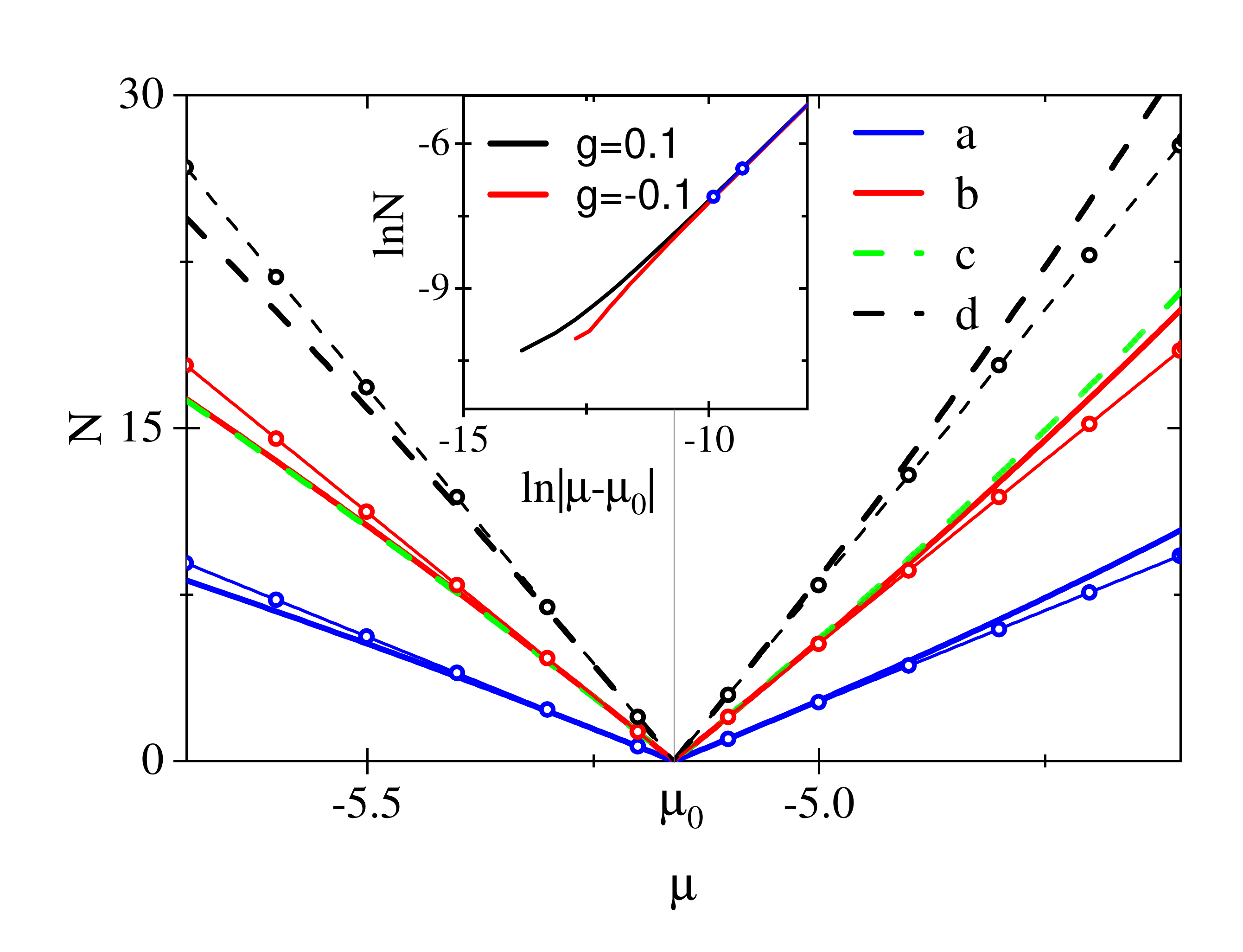}
	\caption{Families of WSs are shown for $g=-0.1$ (in the interval $\mu<\mu_0$) and $g=0.1$ (in the interval $\mu>\mu_0$). The position of the flat band with $\mu_0\approx-5.1608$ and $\Delta_0\approx 2.8\cdot 10^{-6}$, which is not distinguishable on the scale of the plot is indicated by the vertical grey line.   Solid (dashed) lines are used for stable (unstable) families bifurcating at $(\mu, N)=(\mu_0,0)$ from the Bloch mode. Notice that the b and c families are nearly indistinguishable on the scale of the figure. In both panels, lines are for numerical results of Eq.~(\ref{GPdimless}), while the lines with circles are for the results obtained from the approximation Eq.~(\ref{eq_a}). Different colors, also marked by different letters, correspond to the soliton families whose profiles are illustrated in Fig.~\ref{fig_Gapsoliton}. The ln-ln plot in the inset zooms in the region in the vicinity of the flat band $\mu_0$ for one-soliton families bifurcating from the flat band in the limit $N\to 0$ for attractive (the red line) and repulsive (the black line) condensates (the linear band in the inset is located at $-\infty$ of the abscissa).  Here and in all figures below, unless specified otherwise,  we set $V_0=2, \gamma=2.01$, $\Omega=11.4844$ corresponding to point c in Fig.~\ref{fig:one}(a). }
	\label{fig_mu_N}
\end{figure}

The condition (\ref{eps}) corresponds to a relatively wide interval of the system parameters. In particular, it is  valid for the whole range of SOC strength shown in Fig.~\ref{fig:one} (a) and (b).  
In the domain specified by (\ref{eps}) the family is characterized by the linear dependence $N=\chi_0(\mu-\tmu_0)$. (As it is customary, since $\epsilon$ is a formal small parameter, in the final simulations it is set equal to one while $a_0$ is considered to be small).
This is verified in Fig.~\ref{fig_mu_N} using the numerical solution of Eq.~(\ref{GPdimless}) by means of the Newton relaxation as well as the Difference methods. In the numerical calculation, we used the step $dx=\pi/2^{11}$ (verifying the convergence for the smaller steps).  

\begin{figure}[t]
	\includegraphics[ width=0.5\textwidth,trim={1.5cm 0.5cm 2cm 1.5cm}]{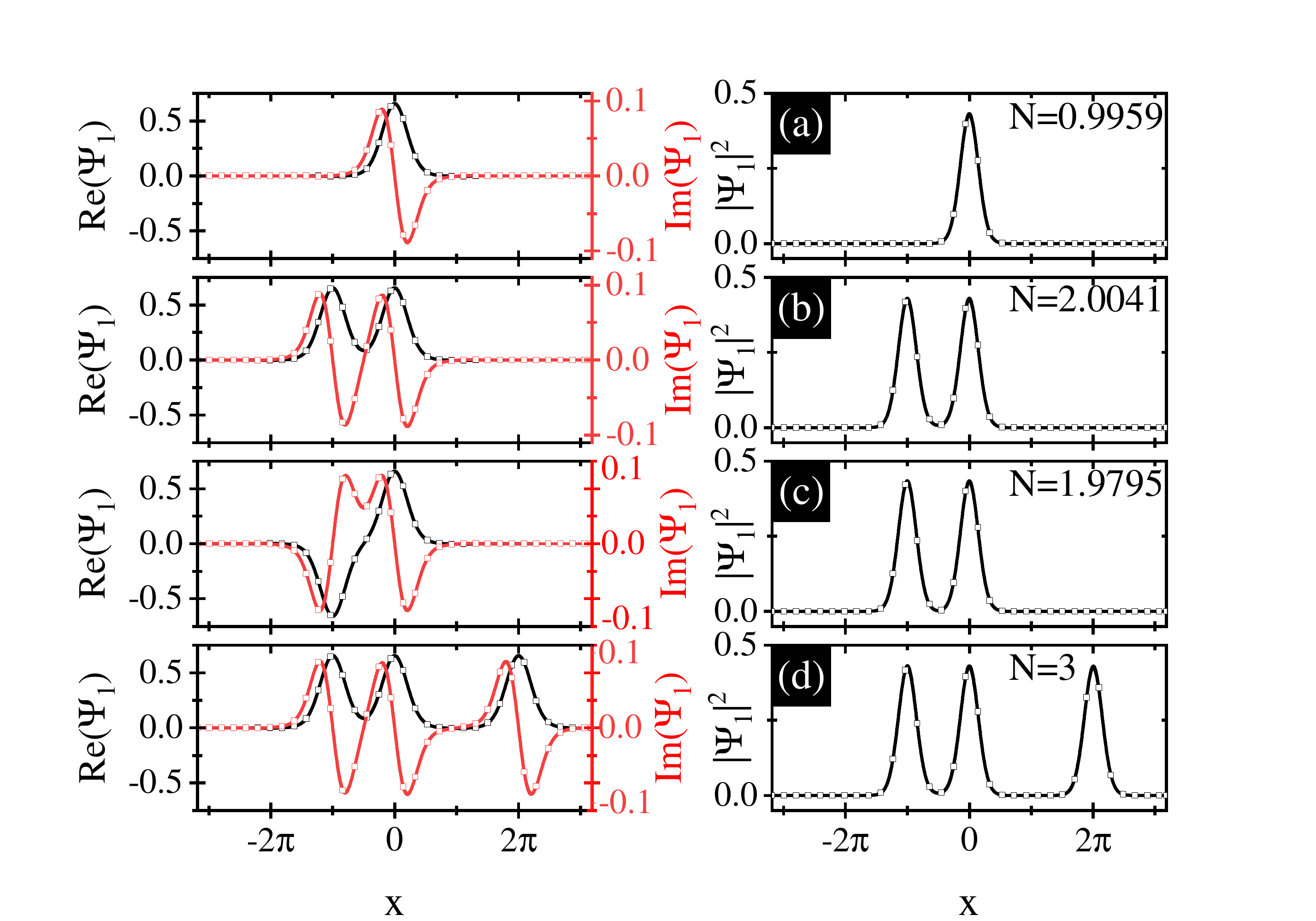}
	\caption{Soliton profiles illustrated for the families indicated in Fig.~\ref{fig_mu_N} by the respective letters, obtained for $g=-0.1$ \rev{arbitrarily chosen value} $\mu=-5.2217$ (solid lines), compared with the linear WFs (dots). In the left (right) column we show the real and imaginary (absolute values) of the first component $\Psi_1$. \rev{Panels (a,b) show fully symmetric solitons while panels (c,d) show solitons with broken symmetry (see the text).}}
	\label{fig_Gapsoliton}
\end{figure}

In Fig.~\ref{fig_mu_N}, the one-soliton [an example is illustrated in Fig.~\ref{fig_Gapsoliton}(a)] families for the attractive $g=-0.1$ ($\mu<\mu_0$)  and repulsive $g=0.1$ ($\mu>\mu_0$) nonlinearities are shown by blue solid lines. \rev{Each family bifurcates from the linear band shown by a thin vertical line in Fig.~\ref{fig_mu_N}. This, however, occurs in the anomalously narrow vicinity of the band, which is a peculiarity of the flatness of the band. Since the mentioned domain is not visible on the scale of the main panel of the figure, in the inset, we zoom this region showing it on the ln-ln plot. Clearly, the linear band on such a plot is located at $-\infty$ of the abscissa, i.e., the bifurcation point itself cannot be shown graphically.}
Beyond the tiny transition region near $N=0$, these families are characterized by the predicted linear dependence $N=\chi_0 (\mu-\mu_0)$ with the slopes $\chi_0>0$ and $\chi_0<0$ for the repulsive and attractive BECs. In the   region close to $\mu=\mu_0$ (zoomed in the inset), the family behaves as $N\sim\sqrt{|\mu-\mu_0|}$ and the solitons become extremely wide at $\mu\to\mu_0$, thus breaking the acceptable numerical accuracy. The 
transition from the gap-soliton interval of the family to the WS linear dependence is confirmed numerically. The observed localization of WSs indicates on the possibility of their experimental observation in low-density condensates, where conventional gap solitons would have extents much larger than the sizes of the typically used trapping potentials.

\begin{figure}[t]
	\includegraphics[ width=0.5\textwidth,trim={0cm 4.5cm 4.5cm 1.3cm}]{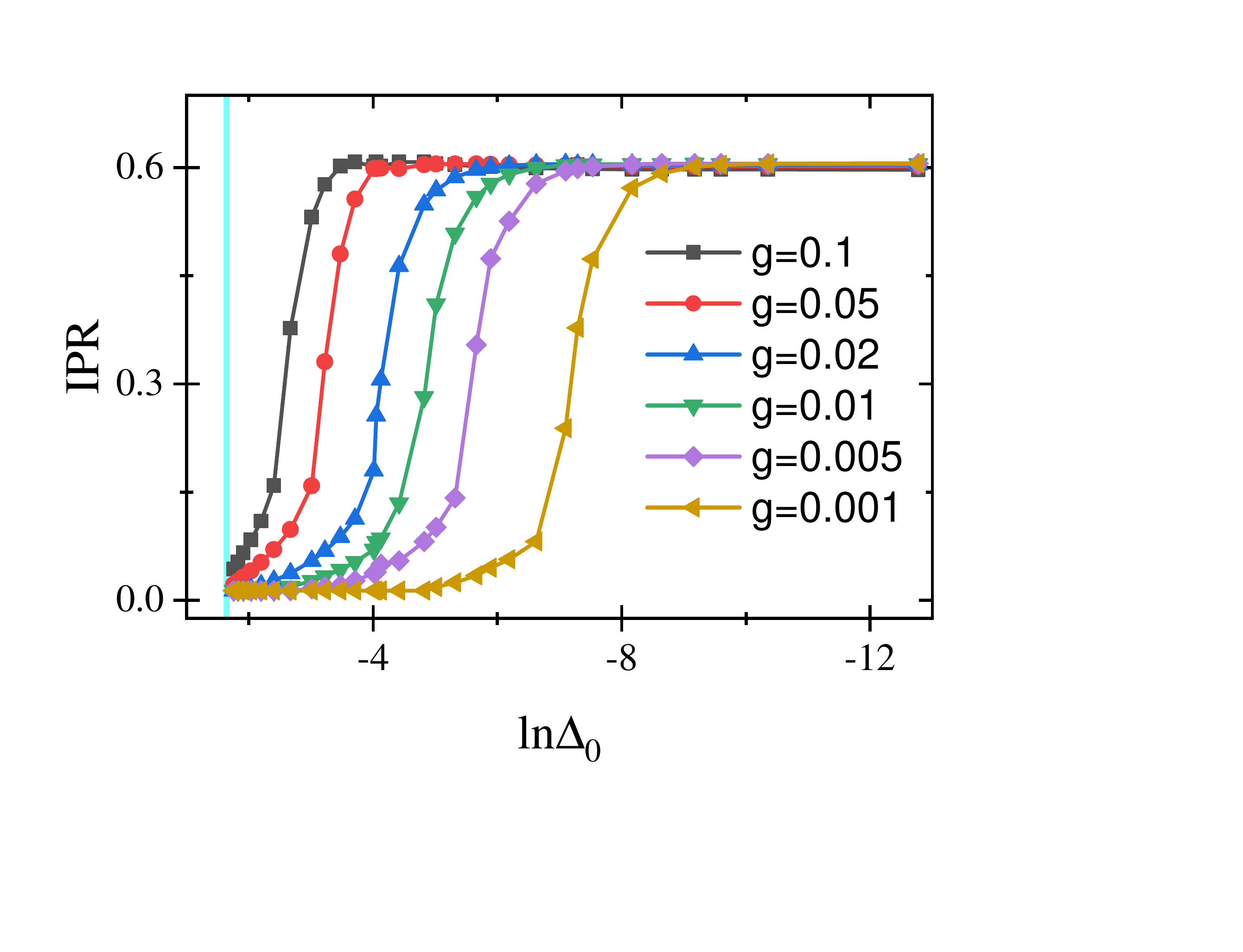}
	\caption{IPR of the fundamental WS {\em versus} ${\ln}\Delta_{0}$ with $\gamma=2.01$ for different non-linearity strengths $g$. Light cyan shadow area corresponds to $\ln\Delta_0\approx -1.758$ corresponding to $\Omega=0$.}	
    \label{fig_IPR}
\end{figure}

A one-peak soliton is confined to a unit cell provided the band is flat.  A numerical illustration of this feature is given in Fig.~\ref{fig_IPR} where we show the localization of WSs characterized by the inverse participation ratio (IPR): 
\begin{align}
 \text{IPR}= \frac{1}{N^2}\int_{-\infty}^{\infty}(\bPsi^\dagger\bPsi)^{2}dx\approx \frac{|\chi_0|}{gN^2} 
\end{align}
  as a function of the flatness parameter $\Delta_0$ (notice the logarithmic scale of the abscissa). To this end, we fix the norm $N=1$ and $\gamma=2.01$ and vary $\Omega$ between 0 and 14.48, i.e., from a conventional non-flat band to (nearly) ideally flat band limits. One can clearly indicate the {\em nonlinear mobility edges} separating areas of the localized and delocalized nonlinear solutions (the conventional linear mobility edge does not exist since our system is periodic). Delocalized gap solitons, IPR$\ll 1 $, near a non-flat band, become well-localized, IPR$\approx 0.6$, above the nonlinear mobility edge even for weak nonlinearity $g=10^{-3}$ (the earthy yellow lines with rotated triangles).  
  
\begin{figure}[t]
	\includegraphics[ width=0.5\textwidth,trim={0cm 4.0cm 0.5cm 1.1cm}]{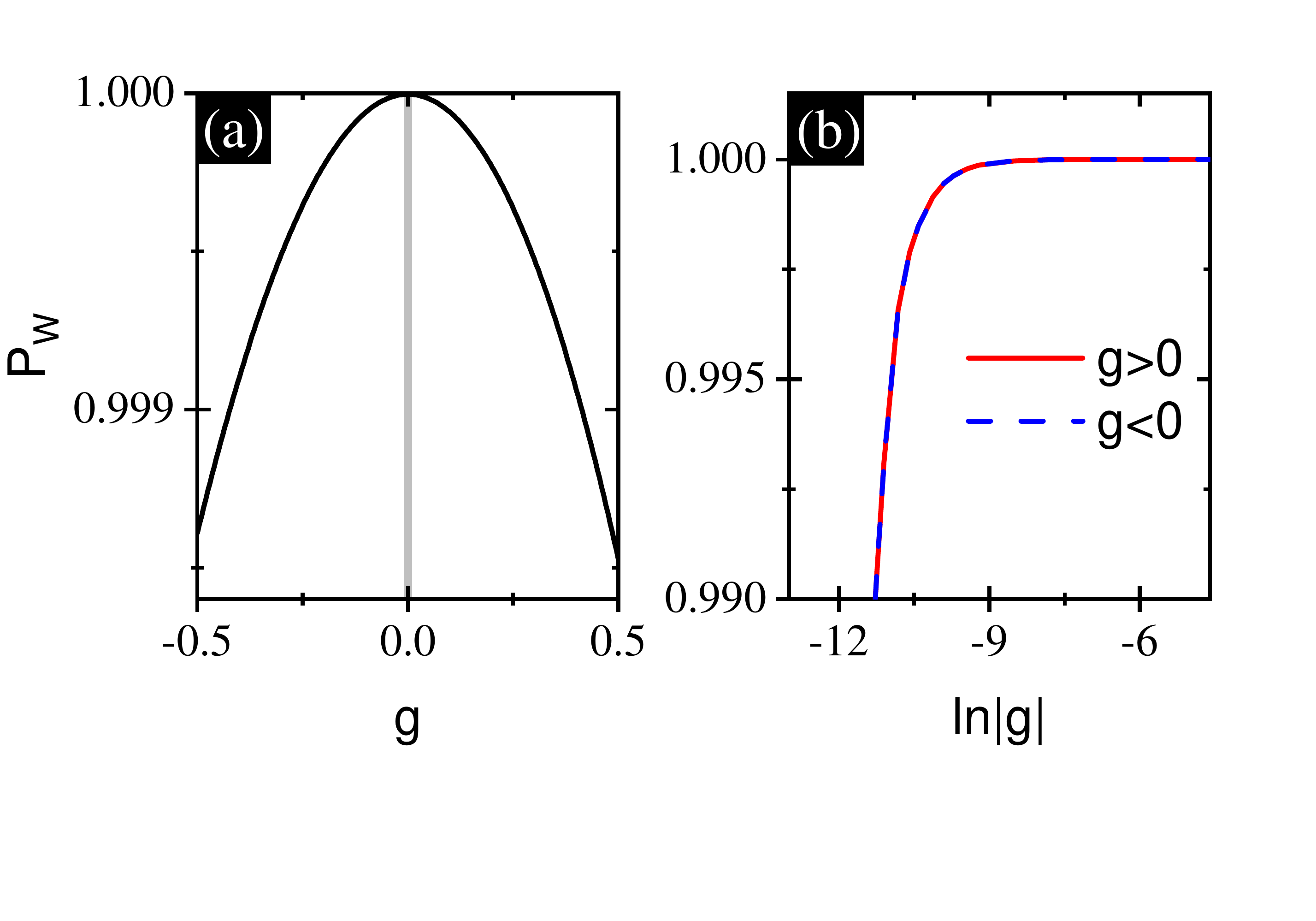}
	\caption{The projection $P_W$ 
 of the fundamental WS on the WF $\bw_0$ 
 {\em vs.} $g$ (a) and its zoom (shown using the logarithmic scale) in the proximity of the flat band (b) for $N=1$. } 
    \label{fig_projection}
\end{figure}

  The widths of WSs remain practically unchanged in a wide interval of the chemical potentials or nonlinearity strengths.  In Fig.~\ref{fig_projection} we plot the projection of a WS on the WF $\bw_0$: $P_W=\langle \bw_{0},\bpsi\rangle$  {\em versus} the nonlinearity coefficient $g$ (for the fixed norm $N=1$). Since the states are strongly localized in the vicinity of a local minimum of the potential, the quantity $1-P_W$ characterizes the population of the upper non-flat bands. We observe that in a sufficiently large interval of magnitude $g$, the projection $P_W$ is nearly one [Fig.~\ref{fig_projection} (a)] and it rapidly decreases in the close vicinity of the flat band [Fig.~\ref{fig_projection} (b)]  where the expansion (\ref{expansion}) is not applicable. 
 
We have studied both, the linear stability of WSs by solving the standard Bogoliubov--de Gennes equations, as well as the (in)stability obtained by simulating their long-time behavior within Eq.~(\ref{GPdimless}) with the addition of initial Gaussian noise of the order of  10\%  of the soliton profile. Both approaches manifested fully consistent results.
The results of this analysis are summarized in Fig.~\ref{fig_mu_N} where stable (unstable) families are shown by solid (dashed lines). The family of one-hump WSs is found stable in the whole region of the parameters where the approximation $N\approx\chi_0(\mu-\mu_0)$ holds.

In addition to the discrete translation symmetry the Hamiltonian  (\ref{H}), and thus the GPE (\ref{GPdimless}), has a Klein four--group of symmetries $\{1,\halpha_1,\halpha_2,\halpha_3\}$, where $\halpha_1=-\mathcal{P}\sigma_x$, $\halpha_2=-\mathcal{K}\sigma_x$, $\halpha_3= \mathcal{P}\mathcal{K}$, $\mathcal{P}$ is the inversion operator, and $\mathcal{K}$ is the complex conjugation operator. The shown one-hump WSs obey all mentioned symmetries. Meantime, the strong localization of WSs suggests the existence of multi-hump solitons, which loosely can be understood as bound states of the fundamental WSs and may obey \rev{either all or} only one of the symmetries $\halpha_j$.
Three examples are shown in Fig.~\ref{fig_Gapsoliton}(b-d). 
A two-hump WS belonging to the fully-symmetric soliton family, i.e., obeying the mentioned Klein four-group symmetries, is shown in Fig.~\ref{fig_Gapsoliton}(b). The soliton shown in Fig.~\ref{fig_Gapsoliton}(c) obeys only $\halpha_1$ symmetry, thus representing {\em nonlinear symmetry breaking} (solitons belonging to this family are anti--$\halpha_2$ and anti--$\halpha_3$ symmetric).   
Respectively, these families are characterized by nearly coinciding dependencies $N(\mu)$ (cf. red solid line b and green dashed line c in Fig.~\ref{fig_mu_N} where they are practically indistinguishable). Every two-hump WS carries nearly double the number of atoms compared with the one-hump soliton at the same value of the chemical potential. 
We have found that the fully-symmetric branch b, characterized by the linear dependence $N\sim\mu$, is stable. Meantime, the $\halpha_1$-symmetric branch c is unstable over the whole domain of chemical potentials shown in Fig.~\ref{fig_mu_N} (except the very narrow region in the vicinity of $\mu_0$ shown in the inset, which was not investigated because of numerical constraints).  

Finally, in Fig.~\ref{fig_Gapsoliton}(d) we illustrate a three-hump WS obeying only $\halpha_3$ symmetry ($\halpha_2$ and $\halpha_3$ involving  inversion $\mathcal{P}$ being broken). Such solitons existing at relatively weak nonlinearities are enabled by flat bands, but we found them all to be unstable (see the respective branch in Fig.~\ref{fig_mu_N}). \rev{Generally, the obtained WSs emerging as strongly loclaized extra objects on the top of the periodic OL contrast with gap solitons in the non-flat band whose internal structure inherits the lattice periodicity}~\cite{Zhang2015}.

\section{Wannier breathers}
\label{breather}

\subsection{A few mode approximation}

The enhanced stability of the WSs as well as their existence in the "quasi-linear" limit determined by the relations (\ref{eps}), suggests a way of constructing oscillatory solutions. Indeed, "trains" of WFs, similar to those shown in Fig.~\ref{fig_Gapsoliton} (b)-(d),  can be viewed as superpositions of single WSs. If the amplitudes of the humps in such a train are weakly modulated, i.e., the train is not an exact nonlinear solution, it is no longer a stationary solution too, and the atomic transfer between the localized clouds can occur. In the case of two coupled WSs such a setting is alike to a BJJ~\cite{BJJ1,BJJ2,BJJ3} and especially to the realization of a BJJ in a quasi-periodic potential reported recently in~\cite{Prates2022}. There is however an essential difference between the results reported below with the previous studies of BJJ: due to the orthogonality of the WFs, the coupling in our model is exclusively nonlinear (it vanishes in the linear limit), i.e., the reported oscillating objects are rather {\em Wannier breathers} than linearly coupled atomic clouds.

Nevertheless, the analogy with BJJ suggests that the theoretical approach for the description of the Wannier breathers can use replication of the general approach elaborated in the mean-field theory of BJJ~\cite{Smerzi1997}. Limiting the consideration to the cases of two- and three-hump breathers with humps in the nearest neighbor potential minima (the generalizations are straightforward), we look for the leading order of a solution of the GPE (\ref{GPdimless}) in the form of the three-mode ansatz  
 \begin{align}
 \bPsi=e^{-i\tmu t}\sum_{n=-1}^1A_{n}(t)\bw_{n}(x).
 \end{align}
Here $A_{n}$ is time-dependent amplitudes of the WF $\bw_{n}$ and excitation of upper bands is neglected. This implies the conservation of the norm $\left|A_{-1}\right|^2+\left|A_{0}\right|^2+\left|A_{1}\right|^2=N$. The dynamical equations for the time-dependent amplitudes are obtained in a standard way (similar calculations for four coupled modes can be found say in~\cite{Prates2022}). 
In  terms  of canonically conjugate variables, which are the  phase  differences $\phi_{\pm}=\arg A_{0}-\arg A_{\pm1}$ and the fractions of the total number of atoms $N_{\pm}={\left|A_{\pm 1}\right|^2}/N$, these equations are obtained from the Hamiltonian
\begin{align}
\label{H_multimode}
    H=&-\frac{\chi_0}{2}
    \left[N_{+}^{2}+N_{-}^{2}+\left(1-N_{+}-N_{-}\right)^{2}\right]\notag\\
    &-2\chi_1
    \sqrt{1-N_{+}-N_{-}}\left[\left(1-N_{-}\right)\sqrt{N_{+}}\cos\phi_{+}\right.\notag\\
    &\left. 
    +\left(1-N_{+}\right)\sqrt{N_{-}}\cos\phi_{-}\right].
\end{align}
and read 
\begin{align}
\label{3modeN}
    \frac{d N_{\pm}}{d\tau}=&2\chi_1
    \left(1-N_{\mp}\right)\sqrt{N_{0}N_{\pm}}\sin\phi_{\pm}\,,
    \\
    \label{3modePhi}
    \frac{d\phi_{\pm}}{d\tau}=&
    \left(N_{\mp}-N_{0}\right)\left(\chi_0-\chi_1\frac{1-N_{\mp}}{\sqrt{N_{0}N_{\pm}}}\cos\phi_{\pm}\right)
    \notag\\
    &-\chi_1
\left(3\sqrt{N_{0}N_{\mp}}+\frac{N_{\mp}^2}{\sqrt{N_{0}N_{\pm}}}\right)\cos\phi_{\mp},
\end{align}
with $\tau=N t$ and $N_0=1-N_--N_+$. 
We emphasize that the Hamiltonian~(\ref{H_multimode}), having apparent similarity with the three-level models governing atoms in three-well traps~\cite{Eckert2006,Liu2017}, describes essentially different dynamics, because the coupling of BEC clouds in our case is non-linear.

\subsection{Evolution of a two-hump breather.}

First, we consider the dynamics of a two-hump breather, which allows for direct comparison with the standard mean-field model for the BJJ~\cite{Smerzi1997,BJJ3}. To this end, we set $A_1=0$. Now the population imbalance $z$  and the relative phase $\phi$ are given by $z=(\left|A_{-1}\right|^2-\left|A_{0}\right|^2)/(\left|A_{-1}\right|^2+\left|A_{0}\right|^2)$ and $\phi=\arg A_{0}-\arg A_{-1}$. One can obtain the reduced Hamiltonian from Eq.~(\ref{H_multimode}): 
\begin{align}
    H=-2\chi_1\sqrt{1-z^2}\cos \phi -\frac{1}{2}\chi_0z^2,
   \label{H_two_model}
\end{align}
 and dynamical equations
\begin{align}
    \frac{dz}{d\tau}&=2\chi_1\sqrt{1-z^2}\sin \phi,
    \label{eq_two1}
     \quad
    \\
    \frac{d\phi}{d\tau}&=\chi_0z- 2\chi_1\frac{z\cos \phi}{\sqrt{1-z^2}} .
    \label{eq_two2}
\end{align}
These equations are akin to the usual two-mode models, but the coupling and hence the dynamics disappear in the linear limit (recall that $\chi_{0,1}\propto g$). Furthermore, from the explicit expression for the stationary points ($0,0$), ($0,\pm\pi$), and ($\pm\sqrt{1-\Lambda^{2}},\pm\pi$) where  $\Lambda=2\chi_1/\chi_0 $, one can see that a self-trapping solution, i.e., a stationary point with imbalanced populations $|z|=\sqrt{1-\Lambda^{2}}$, is determined by the properties of the linear system and is independent of the nonlinearity since  $\Lambda$ is computed using the linear Wannier functions [see (\ref{chi_n})].  
\begin{figure}[t]
	\includegraphics[ width=0.5\textwidth,trim={0cm 0.7cm -0.8cm -0.3cm}]{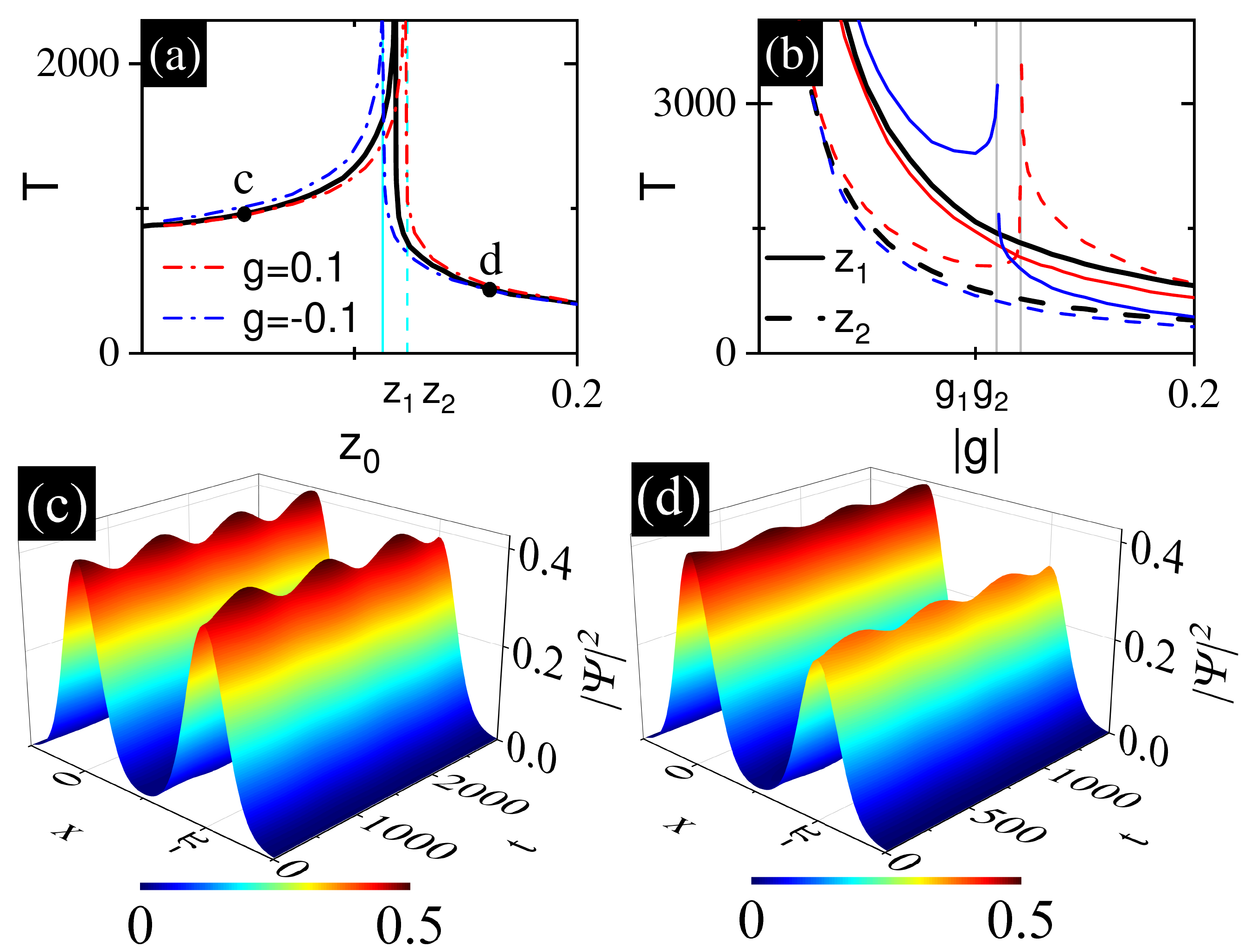}
	\caption{Periods, $T$, of two-hump breathers {\em versus} the initial population imbalance $z_0$ (a) and  {\em versus} interaction strength $g$ (b) for $\phi_{-}=0$, ${\chi_0/g\approx 0.6057}$ and ${\chi_1/g\approx-1.064\cdot 10^{-3}}$. Black lines are analytical results from Eqs.~(\ref{eq_two1}), (\ref{eq_two2}), red (repulsive) and blue (attractive) lines are numerical results from the direct numerical solutions of the GPE (\ref{GPdimless}). The population imbalance corresponding to the separatrix is $\rev{z_*\approx 0.118}$. In panel (b), the grey lines are for transition points, and solid and dashed lines are for $\rev{z_1\approx 0.113<z_*}$ and $\rev{z_2\approx 0.124}>z_*$, respectively, shown in panel (a) with light cyan lines. Two types of evolution at the points indicated in panel (a) are explored in panels (c) (oscillations of the BJJ type) and (d) (oscillations in the self-trapping regime). In both    panels $g=0.1, A_{1}=0, A_{0}=1$, while  $A_{-1}=0.95$ in (c) and $A_{-1}=0.85$ in (d).}
	\label{fig_breather_2}
\end{figure}

The dynamical regimes governed by the three-mode model (\ref{H_multimode})--(\ref{3modePhi}) are insensitive to the sign of interactions (up to the time reversal): the phase portraits governed by Hamiltonian (\ref{H_two_model}) coincide for positive and negative $g$. Each phase portrait has a separatrix, with 
 the imbalance denoted by $z_*$, at which the period of oscillations diverges [see the black solid line in Fig.~\ref{fig_breather_2}(a)].
However, the realistic dynamics, while preserving qualitative separation between "Josephson oscillations", i.e., oscillations around $z=0$ [Fig.~\ref{fig_breather_2}(c)], and oscillations in the vicinity of the "self-trapping" point [Fig.~\ref{fig_breather_2}(d)] manifests differences in the vicinity of the separatrix. These differences are illustrated by blue and red dashed-dotted lines in  Fig.~\ref{fig_breather_2}(a). 

Since the oscillations shown in Figs.~\ref{fig_breather_2}(c) and (d) are due to nonlinear coupling, they disappear in the linear limit. This is expressed in Fig.~\ref{fig_breather_2}(b) by the divergences of all the curves $T(|g|)$ in the limit $g\to 0$. Furthermore, direct numerical simulations show that not all features of the dynamics in the vicinity of the separatrix can be captured by the two-mode model. In Fig.~\ref{fig_breather_2}(b) one observes divergence of the periods of oscillations in the points $g=-g_1$ with initial imbalance $z_1<z_*$ (solid blue line) and $g=g_2$ with initial imbalance $z_2>z_*$ (red dashed line). These singularities of the real dynamics can be explained by the "shift of the separatrix" towards smaller and bigger imbalances $z_1$ and $z_2$, respectively, as shown in Fig.~\ref{fig_breather_2}(a). Thus, when $|g|$ increases at a fixed population imbalance, the separatrix can be crossed either only by the attractive condensate with the imbalance $z_1$ or only by the repulsive condensate with the imbalance $z_2$.  From the physical point of view, the sensitivity of the system in the vicinity of the separatrix can be explained by deviations of the localization, and thus of effective interactions, of the WSs upon the effect of either attractive or repulsive nonlinearity.

Benefited from the enhanced stability of two-hump WSs, two-hump breathers appear to be stable, verified by directly simulating their long-time (up to $t=40000$) behavior with the addition of 5\% Gaussian noise to the initial two-hump profile. We also examined the stability for $|g|=1$ and found that such breathers remain stable having relatively large amplitudes.  

\subsection{On the evolution of the three-hump breather}

\begin{figure}[t]
\includegraphics[ width=0.5\textwidth,trim={0cm 0.7cm -0.8cm -0.3cm}]{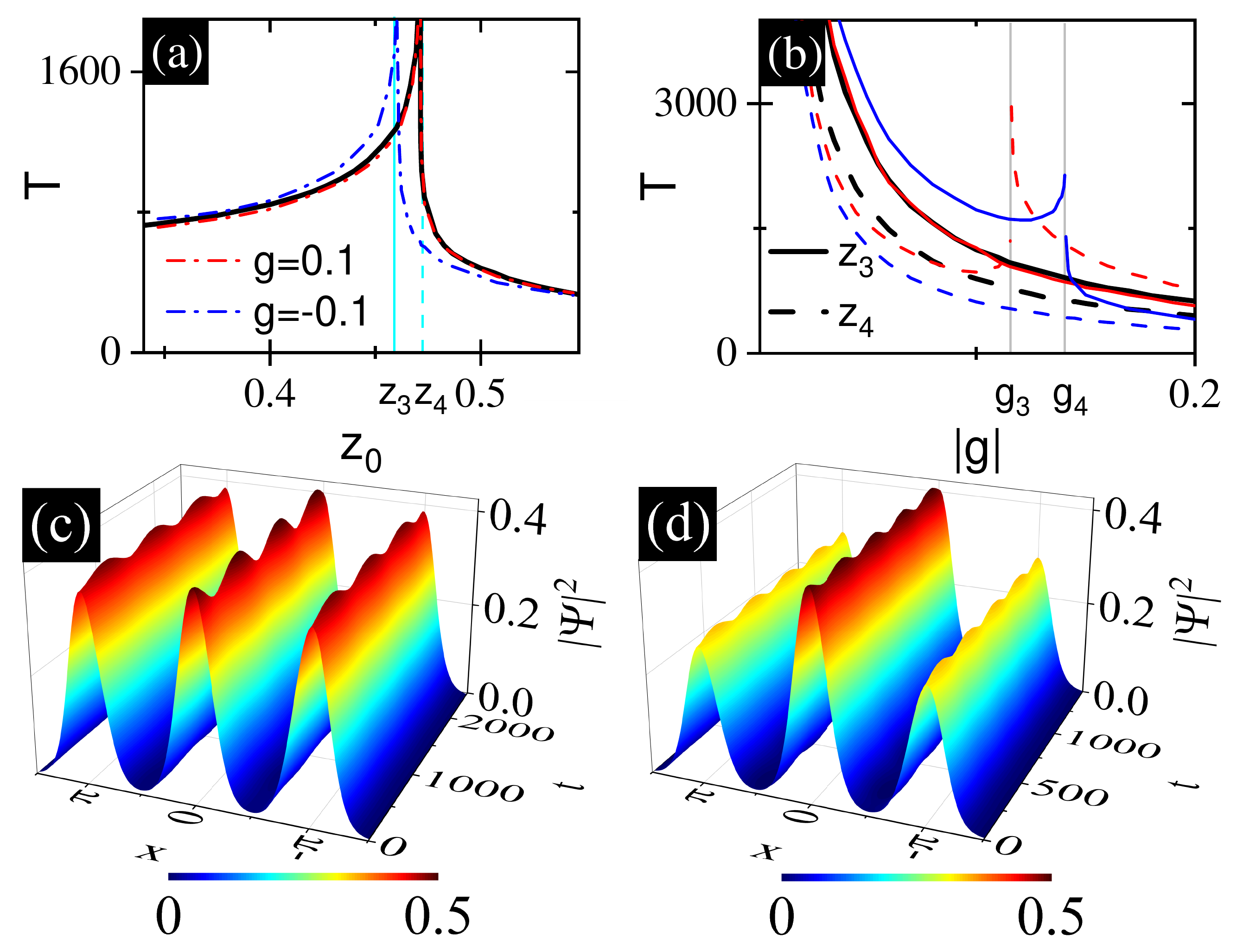}
	\caption{Periods, $T$, of three-hump breathers {\em versus} the initial population imbalance $z_0$ (a) and {\em versus} interaction strength $g$ (b) for $\phi_{\pm}=0$, ${\chi_0/g=0.6067}$ and ${\chi_1/g=-1.064\cdot 10^{-3}}$. Black lines are analytical results from Eq.~(\ref{H_equal}). Red (repulsive) and blue (attractive) lines are numerical results from direct numerical solutions of GPE~(\ref{GPdimless}). The population imbalance corresponding to the separatrix is $\rev{z_*\approx0.471}$. In panel (b), the grey lines are for transition points, and solid and dashed lines are for $\rev{z_3\approx0.459}<z_*$ and $\rev{z_4\approx0.472}>z_*$, respectively [they are shown in panel (a) with light cyan lines]. In panels (c) and (d) two types of evolution corresponding to the points indicated in panel (a) are shown for  $g=0.1, A_{0}=1$, and  $A_{\pm1}=0.95$ in (c) and  $A_{\pm1}=0.8$ (d).}
	\label{fig_breather_3}
\end{figure}

The dynamics of a three-hump breather even within the framework of the three-mode approximation described by Eqs.~(\ref{H_multimode})--(\ref{3modePhi}), where the central mode $\bw_{0}$ is coupled to the two nearest-neighbor modes $\bw_{\pm1}$, can be remarkably complex. Therefore, to illustrate the existence of a stable three-hump breather we address only the simplest case where $N_{+}=N_{-}$, $\phi_{+}=\phi_{-}=\phi$. Within the framework of the mode expansion, this assumption leads to a reduced Hamiltonian that is obtained from Eq.~(\ref{H_multimode})  
\begin{align}
\label{H_equal}
    H=-\chi_{0}(3z^{2}-2z)-4\chi_1 \sqrt{2(1-z^{2})}\left(3-z\right)\cos \phi,
\end{align}
where $z=(2\left|A_{-1}\right|^2-\left|A_{0}\right|^2)/(2\left|A_{-1}\right|^2+\left|A_{0}\right|^2)$  [{\it cf}. (\ref{H_two_model})]. The equations of motion read
\begin{align}
    \frac{dz}{d\tau}&=4\chi_1\sqrt{2(1-z^2)}(3-z)\sin\phi,\\
    \frac{d\phi}{d\tau}&=2\chi_0(3z-1)+\frac{8\chi_1(2z^2-3z-1)}{\sqrt{2(1-z^2)}}\cos\phi.
\end{align}
Unlike in the case of a two-mode breather, now the fixed points  which are found as the roots of the polynomial
\begin{align}
\sqrt{1-z^2}(3z-1)\pm \sqrt{2}\Lambda (2z^2-3z-1) =0,
\end{align}
where "$+$" and "$-$" stand for $\phi=0$ and $\phi=\pm\pi$, do not correspond to equal populations of all three potential minima. In particular, for the choice of the parameters used in Fig.~\ref{fig_breather_3} the four stationary points are found numerically to be approximately $(0.33,0)$, $(0.34,\pm \pi)$, and   $(\pm 1,\pm\pi)$.

In spite of these differences in the Hamiltonians (\ref{H_two_model}) and  (\ref{H_equal}), the latter showing much richer dynamics, as it is seen from the set of the stationary points, in Fig.~\ref{fig_breather_3} (a) and (b) we observe that the three-mode breather share the same peculiarities of evolution as those discussed above in detail for the two-mode case [cf. Fig.~\ref{fig_breather_2} (a) and (b)]. Examples of the numerical study of the respective three-mode dynamics governed by the original model (\ref{GPdimless}) are shown in Fig.~\ref{fig_breather_3}. Now, we again observe enhanced stability of the three-hump breathers in different dynamical regimes, which was verified by considering the evolution for a long time (up to $t=40000$).

\section{Conclusion}
\label{conclusion}

In this work, we described the families of Wanneir solitons that emerge in a mean-field model of the spin-orbit coupled BEC with a nearly flat band. Such families are characterized by the linear dependence of the number of condensed atoms on the detuning of the chemical potential for both attractive and repulsive condensates. Unlike the conventional gap solitons (existing in systems with non-flat bands) Wannier solitons have shapes described by the Wannier functions and widths that remain practically unchanged over the significant domain of variation for the chemical potential. When solitons share the whole symmetries of the Hamiltonian they were found to be stable. This allows for the construction of Wannier breathers which represent nonlinearly-coupled condensate clouds. The evolution of the breather is well captured by few-mode models. Since the coupling in our model is nonlinear, the type of evolution of the breathers is determined by the number of atoms (or alternatively, by the scattering length). Two-- and three--hump breathers, considered as examples were also found stable. The considered system also supports families of Wannire solitons, which obey only one of the symmetries of the nonlinear Hamiltonian, i.e., manifests the nonlinear symmetry breaking. The respective families, however, were found to be unstable.
Finally, similar Wannier solitons can be found in other continuous systems, chiefly in optical nonlinear stratified media,  having flat bands which are well separated from the rest of the spectrum.

\acknowledgments

The work was supported by the Portuguese Foundation for Science and Technology (FCT) under Contracts PTDC/FIS-OUT/3882/2020 and UIDB/00618/2020, the China Scholarship Council (CSC) under the Grant CSC N.202206890002. This work was also supported by the National Natural Science Foundation of China with Grants No.11974235 and
11774219.

\end{document}